\documentclass[preprint]{aastex62}
\usepackage{amsmath,amssymb,CJK,soul}
\usepackage{longtable}
\usepackage{hyperref}
\usepackage{longtable}
\usepackage{threeparttablex} 
\usepackage{lineno}
\usepackage{longtable}
\usepackage{rotating}

\newcommand{\ky}{{1998 KY$_{26}$ }}
\newcommand{\kyns}{{1998 KY$_{26}$}}

\received{--}
\revised{--}
\accepted{--}
\submitjournal{AJ}

\shorttitle{Keck and Gemini characterization of \kyns}
\shortauthors{Bolin et al.}

\begin{document}

\title{Keck and Gemini characterization of \textit{Hayabusa2\#} rendezvous target \ky}

\author[0000-0002-4950-6323]{Bryce T. Bolin}
\affiliation{Eureka Scientific, Oakland, CA 94602, USA}
\correspondingauthor{Bryce Bolin}
\email{bbolin@eurekasci.com}

\author[0000-0002-4223-103X]{Christoffer Fremling}
\affil{Division of Physics, Mathematics and Astronomy, California Institute of Technology, Pasadena, CA 91125, USA}
\affil{Caltech Optical Observatories, California Institute of Technology, Pasadena, CA 91125, USA}

\author[0000-0003-4778-6170]{Matthew Belyakov}
\affiliation{Division of Geological and Planetary Sciences, California Institute of Technology, Pasadena, CA 91125,USA}

\author[0000-0003-4863-5577]{Jin Beniyama}
\affiliation{Universit\'{e} C\^{o}te d'Azur, CNRS, Lagrange, Observatoire de la C\^{o}te d'Azur, F 06304, Nice, France}
\affiliation{Institute of Astronomy, Graduate School of Science, The University of Tokyo, Mitaka, Tokyo, 181-0015, Japan}

\author[0000-0002-8963-2404]{Marco Delbo}
\affiliation{Universit\'{e} C\^{o}te d'Azur, CNRS, Lagrange, Observatoire de la C\^{o}te d'Azur, F 06304, Nice, France}
\affiliation{University of Leicester, School of Physics and Astronomy, Leicester, LE1 7RH, UK}

\author[0000-0001-7830-028X]{Robert Jedicke}
\affiliation{Institute for Astronomy, University of Hawai`i at M\={a}noa, Honolulu, HI, 96822}

\author[0000-0001-9665-8429]{Ian Wong}
\affiliation{Space Telescope Science Institute, Baltimore, MD 21218, USA}

\author[0009-0000-8709-5273]{Laura-May Abron}
\affiliation{Griffith Observatory, Los Angeles, CA 90027}

\author[0009-0000-8709-5273]{Keith S. Noll}
\affiliation{Goddard Space Flight Center, Greenbelt, MD 20771, USA}

\author[0000-0002-4434-2307]{Andrew W. Stephens}
\affil{Gemini Observatory/NSF NOIRLab, Hilo, HI, 96720, USA}




\begin{abstract}
Near-earth object (NEO) \ky is a target of the \textit{Hayabusa2\#} spacecraft, which it will rendezvous with in July 2031. The asteroid is a rapid rotator and has a large out-of-plane nongravitational acceleration. We present deep $g$ and $R$ band imaging obtained with the Keck I/Low Resolution Imaging Spectrometer and visible spectroscopy from Gemini North/Gemini Multi-Object Spectrograph taken of \ky on 2024 June 8-9 when the asteroid was $\sim$0.037 au from the Earth. The asteroid lacks evidence of a dust coma in the deep images and its spectrum most closely resembles Xe-type asteroids, possessing a spectral slope of 6.71$\pm$0.43 $\%$ 100 nm$^{-1}$, and colors $g$-$r$ = 0.63$\pm$0.03, $r$-$i$ = 0.15$\pm$0.03, $i$-$z$ = 0.05$\pm$0.04, and implies a diameter of $\sim$10 m. From our images, we compute a 3$\sigma$ upper limit on the dust production of \ky of $<$10$^{-5}$ kg s$^{-1}$, $<$10$^{-2}$ kg s$^{-1}$, and $<$10$^{-1}$ kg s$^{-1}$ assuming $\mathrm{\mu}$m, mm, and cm size dust particles. Additionally, we compare the orbit of \ky and large nongravitational parameters asteroids to NEO population models and find that the majority, including \kyns, likely originated from the inner Main Belt, while the second most numerous group originates from the outer Main Belt, followed by a third group originating from the Jupiter Family Comet population. Given its inner Main Belt origin, its Xe-type spectrum, and rapid rotation, we hypothesize that the nongravitational acceleration of \ky may be caused by the shedding of large dust grains from its surface due to its rotation rather than H$_2$O vapor outgassing.
\end{abstract}

\section{Introduction}

With a semi-major axis of $a$, of 1.23 au, eccentricity, $e$, of 0.2, inclination, $i$, of 1.48 degrees, and absolute magnitude, H of 25.7\footnote{taken from JPL Small-Body Database:\url{https://ssd.jpl.nasa.gov/tools/sbdb_lookup.html\#/?sstr=1998ky26}, accessed on 2024 January 13}  the Apollo Near-Earth object (NEO) \ky was first noted for being one of the smallest asteroids studied by ground-based radar \citep[][]{Ostro1999}. The X-band (3.5 cm wavelength) radar detection of \ky by the Goldstone Solar System Radar facility provided a diameter estimate between 20 m and 40 m, making it stand out as a potential target for testing the detection of the thermal recoil Yarkovsky Effect \citep[][]{Vokrouhlicky2000}. The asteroid was also noted for having a 10.7 min rotation period, well below the $\sim$2.2 h critical spin period limit for surface material near the equator to remain gravitationally bound to the asteroid in the absence of cohesive forces \citep[][]{Pravec2000}.

Recently, \ky was selected as a rendezvous target of the \textit{Hayabusa2\#} extended spacecraft mission, which will reach the asteroid in 2031 \citep[][]{Hirabayashi2021}. The rendezvous is expected to last $\sim$1 year and will include close proximity operations to determine the bulk properties of the asteroid's surface as well as characterize its regolith \citep[][]{Kikuchi2023}. The spacecraft is also expected to characterize the dust environment surrounding the asteroid as well as touch down on the asteroid and conduct kinetic experiments with projectile firing \citep[][]{Saiki2020}.

In addition to being observed by radar and a spacecraft mission rendezvous target, \ky has been identified to possess an orbit with large non-gravitational accelerations, exceeding those provided by thermal recoil forces, and in the out-of-plane direction \citep[][]{Farnocchia20232003RM,Seligman2023,Jewitt2024NonGrav}. In some cases, authors refer to these objects as ``dark comets'', although there is no direct evidence yet for their activity being driven by the sublimation of cometary volatiles such as H$_2$O. \ky has radial, transverse, and out-of-plane acceleration components of 1.60$\pm$0.88 $\times$10$^{-10}$ au d$^{-2}$, -1.38$\pm$0.57 $\times$10$^{-13}$ au d$^{-2}$, and 2.70$\pm$0.65 $\times$10$^{-11}$ au d$^{-2}$, with the out-of-plane noted for being particularly large and significant \citep[][]{Seligman2024}. The non-gravitational acceleration of \ky is interpreted by \citet[][]{Seligman2024} as possibly being caused by the outgassing of volatiles such as H$_2$O. The total of all three non-gravitational acceleration components equals to $\sim$1.62 $\times$10$^{-10}$ au d$^{-2}$, implying that \ky has a mass loss rate of $\sim$3 $\times$10$^{-5}$ kg when assuming that the ejected material consists of H$_2$O molecules escaping from the asteroid's surface at 350 m s$^{-1}$. We note that the uncertainties are large on the estimated non-gravitational acceleration components and thus the total acceleration could be much lower than this.

This work discusses the physical characterization of \ky using imaging taken with the Keck/Low Resolution Imaging Spectrometer (LRIS) and spectroscopy taken with the Gemini North/Gemini Multi-Object Spectrograph (GMOS). We use the combination of visible imaging and spectroscopy to study the physical properties of \kyns, an approach previous demonstrated by \citet[][]{Bolin2021LD2,Bolin2022IVO,Bolin2024E3,Bolin2025PT5}. We also compare the orbital elements of \ky and other asteroids showing strong non-gravitational acceleration with near-Earth object (NEO) population models \citep[][]{Granvik2018,Morbidelli2020albedo,Nesvorny2023NEOMOD} to better understand their origins and predicted physical properties.

\section{Observations}
We used the LRIS instrument on the 10.0 m Keck I telescope to observe \ky on 2024 June 08 14:29 under program 2024A$\_$N019 (PI: B.T. Bolin) when the asteroid was at RA = 23 31 00.2, dec = -00 00 18, 0.037 au (14.3 lunar distances) from Earth, a heliocentric distance of 1.012 au, and had a phase angle of 93.3$^{\circ}$. LRIS possesses a blue camera consisting of two 2Kx4K Marconi CCDs with a pixel scale of 0.135\arcsec pixel$^{-1}$ and a red camera consisting of two Lawrence Berkley National Laboratory 2k x 4k fully depleted, high resistivity CCD detectors with the same pixel scale \citep[][]{Oke1995,Rockosi2010}. Both cameras were used with the 2 x 2 binning with two readout amplifiers. The 560 nm dichroic was used in combination with an SDSS-equivalent $g$ filter \citep[$\mathrm{\lambda_{eff}}$ = 467.2 nm, FWHM = 126.3 nm;][]{Gunn1998} in the blue camera and Cousins $R$ \citep[$\mathrm{\lambda_{eff}}$ = 649.2 nm, FWHM = 167.1 nm, ][]{Cousins1976} in the red camera, similar to the observational set up of \citet[][]{Bolin2020CD3,Bolin2023NeptunianTrojans}. 

A set of 30 s exposures was taken in both the blue and red cameras using the $g$ and $R$ filters in the field containing \ky using the sideral tracking rate to provide background stars for photometric calibration and surface brightness profile comparison without the background stars becoming trailed. A series of four 180 s simultaneous exposures in the two cameras immediately followed, tracking the telescope non-sidereally using \kyns's on-sky motion rate of 8.87\arcsec/min. The asteroid was observed at an airmass of $\sim$1.25 with Keck/LRIS. Sky conditions were clear and background stars in the images containing the \ky field had an average FWHM of $\sim$1.08\arcsec. 

The LRIS images were reduced using the LPipe reduction software \citep[][]{Perley2019}. The images were stacked to increase the signal to noise of the detections \citep[e.g.,][]{Whidden2019}. mosaic of the $g$ and  $R$-band image stacks is shown in panels A and B of Fig.~1. The $g$ and  $R$-band photometry measurements were completed using background solar analog stars from the Pan-STARRS1 catalog \citep[][]{Chambers2016} and color transformations described in \citet[][]{Tonry2012} to convert the Pan-STARRS1 catalog magnitudes to their corresponding $g$ and $R$ filter equivalents. The LRIS $g$ and  $R$-band photometry of \ky and standard stars were measured using a 1.62\arcsec~radius circular aperture (6 pixels) and then subtracting from it the median pixel value within a 2.97--4.32\arcsec~(11--16 pixels) radius sky background annulus.

\begin{figure}
\centering
\includegraphics[scale=0.27]{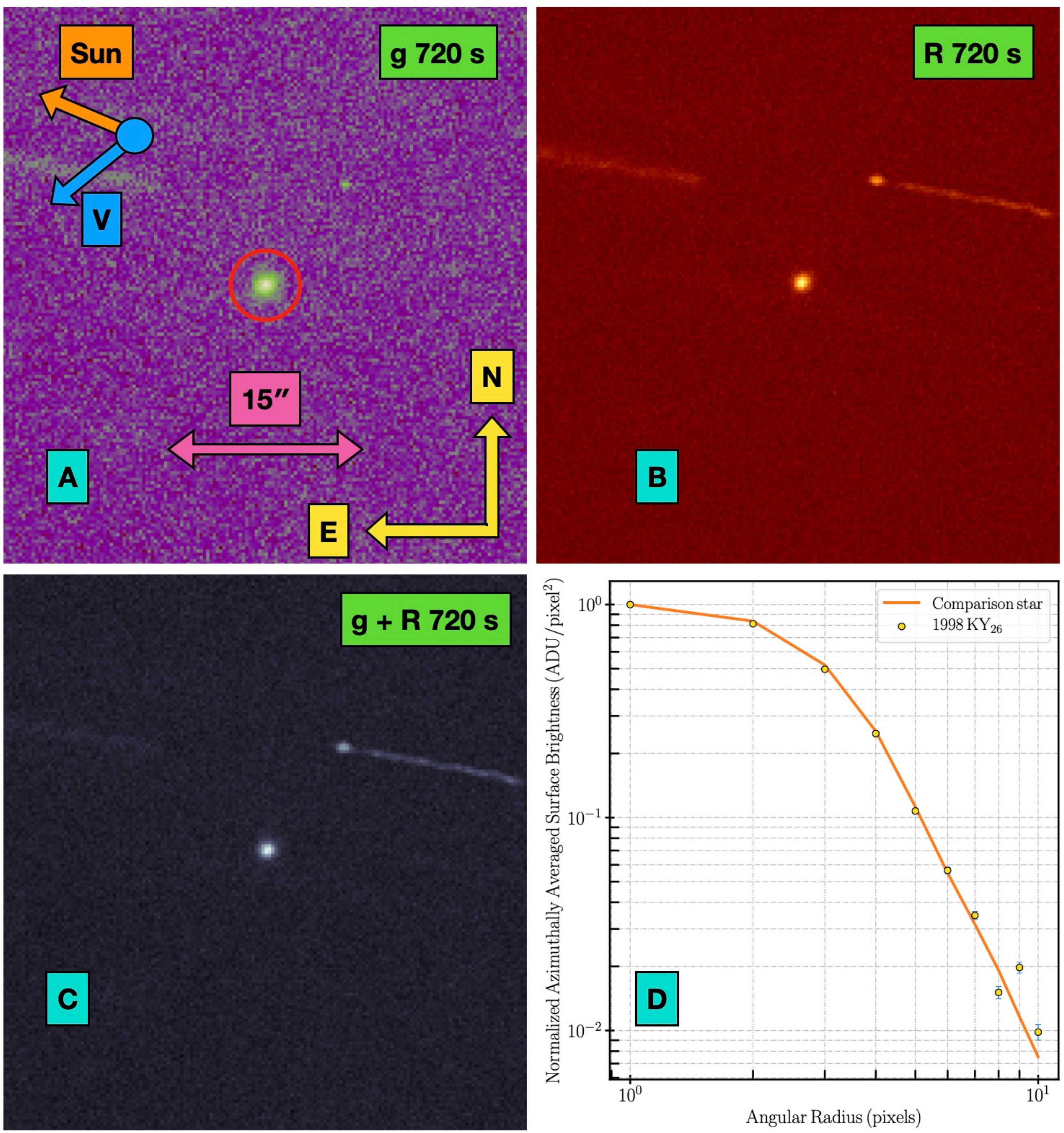}
\caption{\textbf{Deep $g$ and $R$ band imaging of \ky taken with Keck/LRIS on 2024 June 8.} \textbf{Panel A:} a median combination stack of 4 x 180 s $g$ filter LRIS images of \kyns. An arrow indicating a width of 15\arcsec~is shown for scale. The cardinal, solar, and orbital motion directions are indicated. \textbf{Panel B:} a median combination stack of 4 x 180 s $R$ filter images of \kyns. \textbf{Panel C:} a combined stack of all $g$ + $R$ filter images of \kyns. The pixel scale is 0.27\arcsec. Streaks and points are due to incompletely removed background star trails in the image stacks. \textbf{Panel D:} the azimuthally-averaged surface brightness profile of the combined 4 x 180 $g$ + $R$ filter stack plotted as yellow data points with error bars. The error bars are smaller than the marker for the first five data points. A nearby comparison star's azimuthally-averaged surface brightness profile is over-plotted as a piecewise between pixel orange line.}
\end{figure}

The GMOS instrument on the 8.1 m Gemini N telescope was used to observe \ky the following night, on 2024 June 09 13:32, under program GN-2024A-DD-108 (PI: B.T. Bolin) when the asteroid was at RA = 23 45 02.0, dec = +00 37 14, 0.039 au (15.2 lunar distances) from the Earth, a heliocentric distance of 1.011 au, and had a phase angle of 82.1$^{\circ}$. A similar approach was taken to observe \ky with GMOS as previously described by \citet[][]{Bolin2023Dink}. The GMOS detector consists of three 2048 x 4176 Hamamatsu chips with an effective pixel scale of 0.0807\arcsec pixel$^{-1}$. The detectors were rebinned in both the spatial and spectral directions by a factor of two. The R400 grating was used with the 1.5\arcsec slit, the GG455$\_$G0305 order-blocking filter, and a central wavelength of 700 nm to provide spectral coverage between 450 and 1030 nm at an average effective spectral resolution of 0.15 nm \citep[][]{Hook2004}. A nearby solar analog star, HD 215428, was observed for telluric and slope correction using the same GMOS instrument settings. 

A seeing FWHM of $\sim$0.4\arcsec was measured from nearby background stars, and the observations were taken at the local parallactic angle for both the observations of \ky and the solar analog star. The telescope was tracked at the asteroid's sky motion rate of 9.89\arcsec/min, and 6 x 600 s exposures were taken, and the asteroid was observed at airmass of $\sim$1.5. In two of the six exposures, \kyns's spectral trace was contaminated by background stars crossing its path; these exposures were excluded from the analysis, leaving an effective total integration time of 2400 s. The DRAGONS Gemini reduction software \citep[][]{Labrie2023} and a custom pipeline were used to reduce the \ky and solar analog spectra.

\section{Results}

We measured brightnesses of $g$ = 21.57 $\pm$ 0.03, $R$ = 20.72 $\pm$ 0.03 mag from the 720 s LRIS images. In addition, we stacked the $g$ and $R$ band images into a composite $g$ + $R$ stack, as seen in Panel C of Fig.~1, to search for evidence of an extended coma from dust production in \kyns. From the $g$ + $R$ target and comparison star composite stacks, we measured the azimuthally averaged surface brightness profile of \ky and nearby comparison stars, as shown in Panel D of Fig.~1. The surface brightness profile of \ky does not show any evidence of an extended coma when compared to the surface brightness profile of nearby stars.

We computed the GMOS reflectance spectrum of \ky by dividing the asteroid spectrum by the solar analog spectrum and normalizing to unity at 550 nm, as shown in red in Fig.~2. We rebinned the spectrum by a factor of 70 to increase the signal-to-noise ratio. We do not show the data shortward of 550 nm and longward of 930 nm due to the sharp decrease in the signal-to-noise ratio of the spectrum in these wavelength ranges. 

\begin{figure}
\centering
\includegraphics[scale=.4]{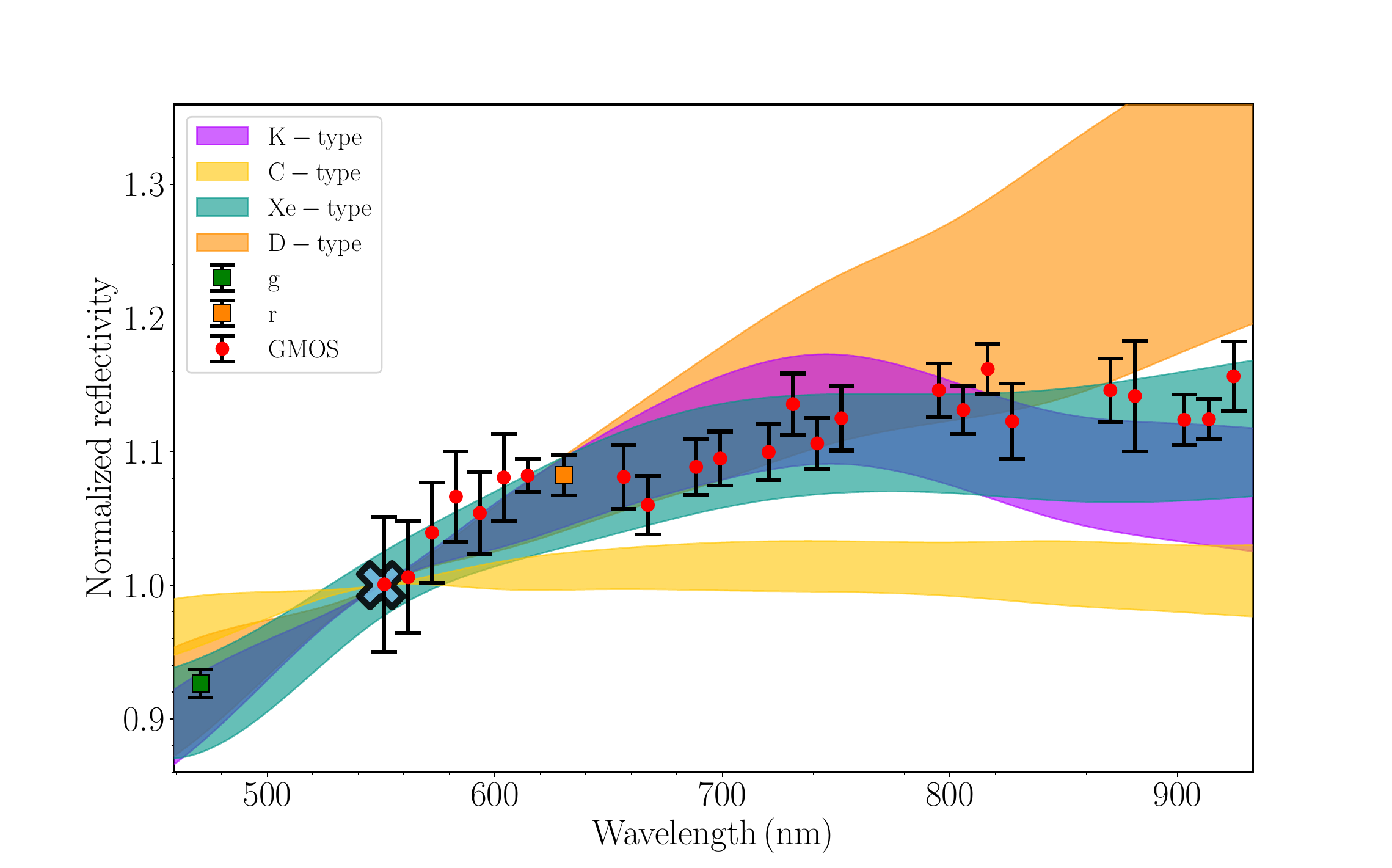}
\caption{\textbf{Visible spectra of \ky taken with Gemini N/GMOS on 2024 June 9.} The Gemini N/GMOS spectrum of \ky is plotted in red, covering 550 nm to 920 nm and normalized to unity at 550 nm (indicated by the light blue cross). The normalized reflectivity from the Keck/LRIS $g$ and $R$ filter \citep[converted to $r$,][]{Jordi2006} observations are included in green and orange, respectively. The gaps in the spectrum at 640, 780, and 850 nm are due to incomplete removal of telluric features. The data are rebinned by a factor of 70 using an error-weighted mean. The spectral ranges of K-, C-, Xe-, and D-type asteroids from \citet[][]{DeMeo2009} are overplotted in purple, gold, aqua, and orange.}
\end{figure}

We incorporate the LRIS $g$- and $R$-band photometry into the combined GMOS + LRIS reflectance spectrum of \ky by dividing the equivalent band fluxes by the corresponding flux of a solar analog and normalizing them to a wavelength of 550 nm. The normalized reflectivity of \ky in the $g$ and $R$ bands are plotted as green and orange data points in Fig.~2, respectively. We use the combined GMOS + LRIS spectrum to compute the spectral slope between an equivalent SDSS $g$-band wavelength ($\mathrm{\lambda_{eff}}$ = 467.3 nm) and an equivalent SDSS $r$-band wavelength ($\mathrm{\lambda_{eff}}$ = 614.1 nm) to obtain a spectral slope value of 10.48$\pm$0.35 $\%$/ 100 nm, which flattens out to an overall spectral slope of 6.71$\pm$0.43 $\%$/ 100 nm when extending this range redward to the equivalent $i$-band wavelength ($\mathrm{\lambda_{eff}}$ = 745.9 nm). Additionally, we compute color indices between the SDSS $g$, $r$, $i$, and $z$ ($\mathrm{\lambda_{eff}}$ = 892.3 nm) bands: $g-r$ = 0.63$\pm$0.03, $r-i$ = 0.15$\pm$0.03, $i-z$ = 0.05$\pm$0.04. These colors are comparable to the equivalent $g-r$ = 0.60$\pm$0.06 and $r-i$ = 0.21$\pm$0.05 color measurements derived from previously published photometric observations of \ky taken by \citet[][]{Ostro1999}.

We compare the spectrum of \ky with the mean Bus-DeMeo asteroid spectra \citep[][]{Bus2002, DeMeo2009}, finding that it resembles Xe-type spectra by visual inspection. We compute the $\chi^2$ statistic for the spectrum of \ky compared to the spectrum of 19 different asteroid types --- S-complex asteroid types (S, Sq, Sv, Q),  C-complex types (B, C, Cg, Cgh), X-complex types (X, Xc, Xe, Xk, Xn), and other assorted types (A, D, K, L, O, R, V) --- from the Bus-DeMeo asteroid spectral catalog and determine the closest match to be Xe-types, with a reduced $\chi^2$ metric of $\sim$1, followed by K-types with a reduced $\chi^2$ of $\sim$3. By comparison, the reduced $\chi^2$ between the spectrum of \kyns and C-types is $\sim$22, and with D-types is $\sim$14. We estimate the size of \ky with $D \, = \, 1329 \, (10^{-H/5} / \sqrt{p_v}$)  from \citet[][]{Russell1916}, where $D$ is the diameter of an asteroid, $H$, is its absolute magnitude, and $p_v$ is its visual albedo. We assume \ky has $p_v$$\sim$0.5, the same as for  since Xe-type asteroids have been shown to have albedos $\sim$0.5 \citep[][]{Delbo2019Ancient}, which combined with its $H$ of $\sim$25.7 results in a $D$ estimate of $\sim$10 m.

Using the composite spectrum of \ky and the combined$g$ + $R$ band LRIS image stack, we estimate upper limits on the Johnson $V$-band equivalent \citep[$\mathrm{\lambda_{eff}}$ = 550.0 nm, ][]{Johnson1953} surface brightness of faint coma centered on the target. We find a $V$-band equivalent 3$\sigma$ upper limit on the surface brightness in a 1.62\arcsec--2.16\arcsec~annulus of $<$1.01 $\times$ 10$^{-7}$ Jy arcsec$^{-2}$, corresponding to a total flux of $<$6.50 $\times$ 10$^{-7}$ Jy. We estimate the absolute magnitude $H$ of a single grain of dust using $H \, = \, 5 \,\log_{10}\left(1329/D\sqrt{p_v}\right)$. We assume that dust from \ky has $p_v$$\sim$0.5, the same as for  since Xe-type asteroids. For $\mathrm{\mu}$m, mm, and cm-scale dust, we obtain $H$ values of $\sim$60, $\sim$46, and $\sim$41.

The $V$-band brightness of a single grain of dust is found using the phase function from \citet[][]{Bowell1988}:
\begin{equation}
\label{eqn.brightness}
V = H + 5\, \mathrm{log_{10}}(r_h \Delta) +2.5\,\mathrm{log_{10}}\left[ (1 - G)\,\Phi_1(\alpha) + G\,\Phi_2(\alpha) \right ]
\end{equation}
where for the 2024 June 08 observations of \kyns, $r_h$ is its heliocentric distance of 1.012 au, $\Delta$ is its geocentric distance of 0.037 au and $\alpha$ is its phase angle of 93.3$^\circ$. $G$ is the phase coefficient, for which we use the value of 0.2, the average value for X-type asteroids \citep[][]{Veres2015}. $\Phi_1(\alpha)$ and $\Phi_2(\alpha)$ are the basis functions normalized at $\alpha$ = 0$^\circ$ described in \citet{Bowell1988}. 

We obtain $V$ $\simeq$ 57, 42, and 37 mag, or a flux of $\sim$3.8 $\times$ 10$^{-20}$ Jy, $\sim$3.8 $\times$ 10$^{-14}$ Jy, and $\sim$3.8 $\times$ 10$^{-12}$ Jy for $\mathrm{\mu}$m, mm, and cm-scale dust particles, respectively. Dividing the 3$\sigma$ upper limit on the total flux inside the 1.62--2.16\arcsec~annulus of $<$6.50 $\times$ 10$^{-7}$ Jy described above, we obtain 3$\sigma$ upper limits on the dust particle number of $<$10$^{13}$, $<$10$^{7}$, and $<$10$^{5}$ for $\mathrm{\mu}$m, mm, and cm-scale dust, or a total dust mass of $<$10$^{-2}$ kg, $<$10 kg, and $<$100 kg assuming that the dust has the same grain density of Aubrite meteorites: $\sim$3.15 g cm$^{-1}$ \citep[][]{Macke2011}. Aubrites are linked to Xe-types \cite[][]{Cuk2014Auberites,Lucas2019}, suggesting that this is a reasonable estimate for the density of dust produced by \kyns. Dividing the total dust mass 3$\sigma$ upper limits by the integration time of 720 s, we obtain 3$\sigma$ upper limits on the dust production of $<$10$^{-5}$ kg s$^{-1}$, $<$10$^{-2}$ kg s$^{-1}$, and $<$10$^{-1}$ kg s$^{-1}$ for $\mathrm{\mu}$m, mm, and cm-scale dust particles.

\section{Discussion and Conclusions}

Measurement of the non-gravitational parameters of \ky shows that it has a total acceleration of $\sim$1.62 $\times$10$^{-10}$ au d$^{-2}$, which given a 350 m s$^{-1}$ outflow speed implies a total mass loss of $\sim$10$^{-5}$ kg s$^{-1}$ \citep[][]{Seligman2024}. This rate is in line with our 3$\sigma$ upper limit for the production of $\mathrm{\mu}$m-sized dust particles. However, the mass loss scenario described by \citet[][]{Seligman2024} assumes that the mass loss could be in the form of H$_2$O vapor molecules, which would escape the surface of \ky at the speed of sound in H$_2$O vapor, i.e., $\sim$350 m s$^{-1}$ at 1 au from the Sun. We also note the true total acceleration of \ky could be much lower due to the large uncertainties in the estimate of all the non-gravitational acceleration components.

The spectral type and source location of \ky seems to diminish the possibility that it contains volatiles. While some asteroids in the Xe-type group show evidence of hydrated spectral features \citep[][]{Rivkin1995}, it seems less likely that small Xe-types such as \ky would possess H$_2$O vapor given that the majority of Xe-types show little to no evidence of hydration features \citep[][]{Fornasier2011Xtype}. Additionally, meteoritic evidence suggests that Jupiter formed quickly, reaching 20 Earth masses within the first Myr after the formation of solids in the solar system \citep[][]{Kruijer2017JupiterAge}, effectively locking the early ice line from reaching within 3 au from the Sun \citep[][]{Morbidelli2016CondensationLines}. This is outside the range of where \ky is likely to have originated within the Main Belt. Additionally, the lack of hydration amongasteroids in the inner Main Belt is supported by the observation that no active asteroids whose activity has evidence of being driven by the sublimation of volatiles are found in the inner Main Belt \citep[][]{Jewitt2022MBCscometsIII}.

Instead, the mass loss of \kyns, as suggested by its non-gravitational acceleration, may be due to a different mechanism than the outgassing of H$_2$O. The short rotation period of \ky of $\sim$10 minutes, well below the stability limit of $\sim$2 h, suggests that material near its surface may be unstable in the absence of cohesive forces \citep[][]{Pravec2000}. While small asteroids have been observed with rotation periods below 2 h \citep[][]{Warner2009}, small asteroids on the scale of $\sim$100 m have been observed to sporadically eject dust and take on an active appearance as a result of surface mass shedding due to their rapid spins \citep[][]{Purdum2021,Jewitt2022MBCscometsIII}.

Surface material ejected from a rotationally unstable asteroid leaves its surface at speeds comparable to the asteroid's gravitational escape speed \citep[e.g.,][]{Jewitt2014EPPP}. This is also true for the fragments of asteroids that have undergone collisional disruption \citep[e.g.,][]{Bolin2017a,Bolin2017b,Moreno2017}. Using the diameter of 10--30 m estimated from a non-detection of \ky in ground-based mid-infrared observations \citep[][]{Beniyama2025a}, and assuming a density of 2600 kg m$^{-3}$ (i.e., a typical density for Xe-type asteroids; \citealt{Carry2012}), its escape speed is $\sim$0.01 m s$^{-1}$. Rescaling the mass loss of 3 $\times$ 10$^{-5}$ kg s$^{-1}$, assumed for H$_2$O molecules escaping at 350 m s$^{-1}$ \citet[][]{Seligman2024}, to an ejection speed of $\sim$0.01 m  s$^{-1}$ results in a mass loss rate of $\sim$0.1 kg s$^{-1}$, comparable to our 3$\sigma$ upper limits on the production of mm and cm-sized dust derived from our deep imaging observations. Additionally, manuscript by Santana-Ros et al.\footnote{\url{https://www.researchsquare.com/article/rs-5821856/v1}} currently in review at time of writing suggests that that \ky may have an even smaller diameter of $\sim$10 m, similar to our size estimate from Section~3. The $\sim$10 m size estimate from our measurements and the measurements of Santana-Ros would imply a mass loss rate of $\sim$0.1 kg s$^{-1}$ similar to our 3$\sigma$ upper limit on the production of mm-size dust. Additionally, Santana-Ros find a rotation period of $\sim$5 minutes for \ky further strengthening our argument that it is rotating faster than the critical spin limit for surface material to remain gravitational bound to the asteroid. Furthermore, the NEO Bennu was observed by the \textit{OSIRIS-REx} spacecraft to be undergoing mass loss in the form of the ejection of mm- to cm-sized particles from its surface, similar to the size of the particles compatible with our mass loss upper limit estimate for \ky \citep[][]{Hergenrother2020}. \textit{Hayabusa2\#} is designed to characterize the dust environment around \kyns, and will provide a more stringent constraint on its dust production rate when it reaches the asteroid in 2031.

We compare the orbital elements of \ky ($a = 1.23$, $e = 0.20$, $i = 1.48$, and $H = 25.7$) with the \citet[][]{Granvik2018}, \citet[][]{Morbidelli2020albedo}, and NEOMOD \citep[][]{Nesvorny2023NEOMOD,Nesvorny2024NEOMOD3} NEO population models to estimate \kyns's source within the Main Belt and Jupiter Family Comet populations. We find that the most likely source for \ky is the $\nu_6$ resonance located on the inner edge of the inner Main Belt \citep[][]{Milani1994a} with a probability of $\sim$0.67 as seen in Table~1. The following two most probable sources are the Hungarias at 1.9 au and the 3:1 mean motion resonance at 2.5 au, with probabilties of $\sim$0.17 and $\sim$0.16, respectively. Combining these three sources strongly implies that \ky originated from the inner Main Belt inside of 2.5 au from the Sun. We note that Hungarias, a source compatible with the orbit of \kyns, consists mainly of Xe-types \citep[][]{Cuk2014Auberites,Lucas2019}, agreeing with the classification of \ky from our spectroscopic observations.

Table~1 shows a complete list of the NEO source probabilities for all of the objects with large non-gravitational accelerations listed by \citet[][]{Seligman2024}. Sources for some large non-gravitational objects have been described by \citet{Taylor2024origins} and are broadly consistent with our results. We find that the majority of objects more recently described by \citet[][]{Seligman2024} likely originate from the inner Main Belt: 1998 KY$_{26}$, 2005 UY$_{6}$, 2012 UR$_{158}$, 2016 GW$_{221}$, 2013 BA$_{74}$, 2016 NJ$_{33}$, 2013 XY$_{20}$, 2005 VL$_{1}$, 2010 RF$_{12}$, 2010 VL$_{65}$, and 2006 RH$_{120}$. Following our previous discussion of the lack of hydration among asteroids in the inner Main Belt, the likely origin of these asteroids suggests that their non-gravitational acceleration is due to mechanical means, such as the shedding of surface material, as opposed to volatile sublimation. Additionally, the small ($\lesssim$10 m) size of some of these objects, such as minimoon 2006 RH$_{120}$ \citep[][]{Kwiatkowski2009}, may suggest that the non-gravitational accelerations of some of these objects may be due to solar radiation pressure, as has been detected for other 10 m-scale asteroids \citep[e.g.,][]{Bolin2025PT5}.

The second most populous source of large non-gravitational asteroids is located in the outer Main Belt where objects 1998 FR$_{11}$, 2001 ME$_{1}$, and 2003 RM most likely originate, in the vicinity of the 5:2 resonance located at the inner edge of the outer Main Belt at 2.82 au \citep[][]{Todorovic2017}. In addition to being more volatile-rich compared to the inner Main Belt \citep[e.g.,][]{Campins2010Themis,Rivkin2010Themis}, the outer Main Belt is also the location of several asteroid families \citep[][]{Hsieh2018,Xin2024MBC}. The escape of asteroids from the outer Main Belt, where Main Belt comet families are located, into the NEO population suggests that some of the large non-gravitational asteroids escaping from this region of the Main Belt  \citep[][]{Granvik2017} may be retaining volatiles such as H$_2$O, which is driving their non-gravitational accelerations. Two of these objects, 1998 FR$_{11}$ and 2001 ME$_{1}$, also have significantly large Jupiter Family comet (JFC) source probabilities of $\sim$30$\%$, making the JFCs the third most likely source of large non-gravitational objects. This suggests that 1998 FR$_{11}$ and 2001 ME$_{1}$ could be weakly active cometary remnants that originated from the volatile-rich Kuiper Belt \citep[][]{Nesvorny2017shortperiod}. We note, however, that the difference in the JFC source probabilities for some large non-gravitational objects between the \citet[][]{Granvik2018} and \citet[][]{Nesvorny2024NEOMOD3} may be due to differences in assumption in the rate at which JFCs become dormant versus disrupting once they reach perihelion passages below 2.5 au (D. Nesvorn\'{y}, private communication).

\section*{acknowledgments}

The authors wish to thank Alessandro Morbidelli for his help with computing the NEO source probabilities for the objects studied in this manuscript, and Josh Walawender for supporting the Keck/LRIS observations. The authors also wish to thank the Gemini North operations staff for help coordinating the observations described in this manuscript.

Some of the data presented herein were obtained at Keck Observatory, which is a private 501(c)3 non-profit organization operated as a scientific partnership among the California Institute of Technology, the University of California, and the National Aeronautics and Space Administration. The Observatory was made possible by the generous financial support of the W. M. Keck Foundation.

Based on observations obtained at the international Gemini Observatory, a program of NSF NOIRLab, which is managed by the Association of Universities for Research in Astronomy (AURA) under a cooperative agreement with the U.S. National Science Foundation on behalf of the Gemini Observatory partnership: the U.S. National Science Foundation (United States), National Research Council (Canada), Agencia Nacional de Investigaci\'{o}n y Desarrollo (Chile), Ministerio de Ciencia, Tecnolog\'{i}a e Innovaci\'{o}n (Argentina), Minist\'{e}rio da Ci\^{e}ncia, Tecnologia, Inova\c{c}\~{o}es e Comunica\c{c}\~{o}es (Brazil), and Korea Astronomy and Space Science Institute (Republic of Korea).

Keck and Gemini North Observatory is located on Maunakea, land of the K$\mathrm{\bar{a}}$naka Maoli people, and a mountain of considerable cultural, natural, and ecological significance to the indigenous Hawaiian people. The authors wish to acknowledge the importance and reverence of Maunakea and express gratitude for the opportunity to conduct observations from the mountain.

\facility{Keck:I (LRIS), Gemini:Gillett (GMOS)} 

\bibliographystyle{aasjournal}
\bibliography{neobib}

\clearpage
\newpage
\begin{sidewaystable}
\caption{Summary of orbital elements, albedo, diameter, and NEO source probabilities for \ky and other asteroids with large non-gravitational accelerations.}
\centering
\begin{tabular}{lllllllllllllll}
\hline
Object$^{1}$     & $a$$^{2}$    & $e$$^{3}$    & $i$$^{4}$     & $T\mathrm{_J}^{5}$    & $H$$^{6}$     & $p_v$$^{7}$    & $D$$^{8}$      & $p\mathrm{_{Hun}}^{9}$ & $p\mathrm{_{\nu6}}^{10}$ & $p\mathrm{_{Pho}}^{11}$ & $p\mathrm{_{3:1}}^{12}$ & $p\mathrm{_{5:2}}^{13}$ & $p\mathrm{_{2:1}}^{14}$ & $p\mathrm{_{JFC}}^{15}$ \\
           & (au) &      & (deg) &       &       &           & (m)   &      &      &      &      &      &      &      \\ \hline
\textbf{1998 KY$_{26}$}  & 1.23 & 0.20 & 1.48  & 5.19 & 25.74 & 0.23/0.12$^{*}$ & 30$^{2}$ & 0.25/0.08 & 0.68/0.66 & 0.0/0.0  & 0.07/0.24 & 0.0/0.01  & 0.0/0.01  & 0.0/0.0  \\
1998 FR$_{11}$  & 2.81 & 0.71 & 6.67  & 2.89  & 16.42 & 0.1/0.02$^{\dagger}$  & 4900$^{1}$    & 0.0/0.0  & 0.01/0.0 & 0.0/0.0  & 0.06/0.05 & 0.61/0.83 & 0.01/0.10 & 0.30/0.02 \\
2001 ME$_{1}$   & 2.63 & 0.87 & 5.96  & 2.67 & 16.52 & 0.1/0.02$^{\dagger}$  & 4600$^{1}$     & 0.0/0.0  & 0.05/0.02& 0.0/0.0  & 0.18/0.25 & 0.45/0.50 & 0.04/0.14 & 0.28/0.10 \\
2005 UY$_{6}$   & 2.25 & 0.87 & 12.21 & 2.94 & 18.14 & 0.20      & 700 & 0.01/0.01 & 0.59/0.23 & 0.01/0.01 & 0.25/0.62 & 0.13/0.10 & 0.0/0.02  & 0.0/0.0  \\
2003 RM    & 2.92 & 0.60 & 10.85 & 2.96 & 19.64 & 0.14      & 400 & 0.0/0.0  & 0.0/0.0  & 0.0/0.0  & 0.03/0.03 & 0.70/0.40 & 0.12/0.52 & 0.14/0.05 \\
2012 UR$_{158}$ & 2.23 & 0.85 & 3.19  & 3.00 & 20.67 & 0.21/0.02$^{\dagger}$ & 630$^{1}$  & 0.01/0.02 & 0.73/0.38 & 0.0/0.0 & 0.22/0.51 & 0.03/0.07 & 0.0/0.02  & 0.01/0.0 \\
2016 GW$_{221}$ & 0.83 & 0.27 & 3.76  & 7.05 & 24.76 & 0.23      & 31 & 0.22/0.07 & 0.72/0.67 & 0.0/0.0  & 0.05/0.25 & 0.0/0.01  & 0.0/0.0  & 0.0/0.0  \\
2013 BA$_{74}$  & 1.75 & 0.44 & 5.22  & 4.00 & 25.40 & 0.22      & 24 & 0.12/0.05 & 0.80/0.75 & 0.0/0.0  & 0.08/0.19 & 0.0/0.01  & 0.0/0.0  & 0.0/0.0  \\
2016 NJ$_{33}$  & 1.31 & 0.21 & 6.62  & 4.94 & 25.53 & 0.23      & 22 & 0.31/0.10 & 0.59/0.71 & 0.0/0.0  & 0.11/0.19 & 0.0/0.01  & 0.0/0.0  & 0.0/0.0  \\
2013 XY$_{20}$  & 1.13 & 0.11 & 2.85  & 5.52 & 25.64 & 0.23      & 21 & 0.28/0.08 & 0.70/0.36 & 0.0/0.0  & 0.02/0.55 & 0.0/0.0  & 0.0/0.0  & 0.0/0.0  \\
2005 VL$_{1}$   & 0.89 & 0.22 & 0.24  & 6.65 & 26.45 & 0.23      & 14 & 0.22/0.07 & 0.68/0.71 & 0.0/0.0  & 0.10/0.22 & 0.0/0.01  & 0.0/0.0  & 0.0/0.0  \\
2010 RF$_{12}$  & 1.06 & 0.19 & 0.88  & 5.79 & 28.42 & 0.23      & 5.7 & 0.21/0.09 & 0.74/0.76 & 0.0/0.0  & 0.05/0.15 & 0.0/0.01  & 0.03/0.0  & 0.0/0.0  \\
2010 VL$_{65}$  & 1.07 & 0.15 & 4.23  & 5.75 & 29.22 & 0.23      & 4 & 0.22/0.09 & 0.71/0.75 & 0.0/0.0  & 0.07/0.16 & 0.0/0.01  & 0.0/0.0  & 0.0/0.0  \\
2006 RH$_{120}$ & 1.03 & 0.02 & 0.59  & 5.93 & 29.50 & 0.23      & 3.5 & 0.27/0.05 & 0.73/0.82 & 0.0/0.0  & 0.0/0.13  & 0.0/0.0  & 0.0/0.0  & 0.0/0.0 
\end{tabular}
\begin{tablenotes}
\item \textbf{Notes.} (1) object name, (2) semi-major axis, (3) eccentricity, (4) inclination, (5) Tisserand's parameter, where $T\mathrm{_J} \, = \, \frac{a_J}{a} + 2 \, \mathrm{cos} \, i \sqrt{\frac{a}{a_J} (1 - e^2)}$ and $a_J$ = 5.2 au is the semi-major axis of Jupiter, (6) absolute magnitude, (7) visible geometric albedo inferred from the albedo model of \citet[][]{Morbidelli2020albedo} and other sources as noted, (8) object diameter, calculated from the absolute magnitudes in column 6 and the albedos in column 7, as well as diameter measurements from other sources where noted. Source probabilities inferred from \citet[][]{Granvik2018} and NEOMOD \citep[][]{Nesvorny2023NEO}: (9) $\nu_6$ resonance source probability, (10) 3:1 resonance source probability, (11) 5:2 resonance source probability, (12) 7:3 resonance source probability, (13) 8:3 resonance source probability, (14) 9:4 resonance source probability, (15) 11:5 resonance source probability, (16) 2:1 resonance source probability, (17) weak inner resonances source probability, (18) Hungaria source probability, (19) Phocaea source probability, (20) Jupiter family comet source probability. The orbital elements of these asteroids were taken from JPL HORIZONS accessed on 2025 January 13. The value of p$\mathrm{_{3:1}}$ determined from NEOMOD was calculated by combining the NEO source probability for the weak inner Main Belt resonances and the 3:1 mean motion resonance. The value of p$\mathrm{_{5:2}}$ determined from NEOMOD was calculated by combing the probability for NEOs originating from the 7:3, 5:2, and 8:3 mean motion resonances. The value of p$\mathrm{_{2:1}}$ determined from NEOMOD was calculated by combing the probability for NEOs originating from the 9:4, 11:5, and 2:1 mean motion resonances. ($^{\dagger}$) albedos for 1998 FR$_{11}$, 2001 ME$_{1}$, and 2012 UR$_{158}$ derived by combining the diameter measured in NEOWISE observations and their absolute magnitudes. ($^*$) albedo for for 1998 KY$_{26}$ derived by combining the diameter measured by radar observations \citep[][]{Ostro1999} and its absolute magnitude.  $^1$Diameters for 1998 FR$_{11}$, 2001 ME$_{1}$, and 2012 UR$_{158}$ measured from NEOWISE observations. $^2$Diameter for1998 KY$_{26}$ measured from radar observations \citep[][]{Ostro1999}.
\end{tablenotes}
\end{sidewaystable}

\end{document}